# Spin-flop led peculiar behavior of temperature-dependent anomalous Hall effect in Hf/Gd-Fe-Co

Ramesh Chandra Bhatt,[1,2] Lin-Xiu Ye,[1,2] Ngo Trong Hai,[3] Jong-Ching Wu,[3] and Te-ho Wu,[1,2,a]

[1]Graduate School of Materials Science, National Yunlin University of Science and Technology, Douliu, Yunlin 64002 Taiwan ROC

[2]Taiwan SPIN Research Center, National Yunlin University of Science and Technology, Douliu, Yunlin 64002, Taiwan ROC

[3]Department of Physics, National Changhua University of Education, Changhua 500, Taiwan ROC

[a]Corresponding author: wuth@yuntech.edu.tw

**ABSTRACT**

Here we investigate the temperature dependence of anomalous Hall effect in Hf/GdFeCo/MgO sheet film and Hall bar device. The magnetic compensation temperature ($T_{comp}$) for the sheet film and device is found to be ~240 K and ~118 K, respectively. In sheet film, spin-flopping is witnessed at a considerably lower field, 0.6 T, close to $T_{comp}$. The AHE hysteresis loops in the sheet film have a single loop whereas in the Hall bar device, hystereses consist of triple loops are observed just above the $T_{comp}$. Moreover, the temperature-dependent anomalous Hall resistance ($R_{AHE}$) responds unusually when a perpendicular magnetic field is applied while recording the $R_{AHE}$. The zero-field $R_{AHE}$ scan suggests the Hall signal generates solely from the FeCo moment. However, the behavior of 3 T-field $R_{AHE}$ scan in which the $R_{AHE}$ drops close to zero near the $T_{comp}$ seems to be following the net magnetization response of the device, is explained by considering the low field spin-flopping around the compensation temperature. The results presented here give important insight to understand the complex AHE behavior of ferrimagnets for their spintronic applications.

**INTRODUCTION**

The extraordinary Hall effect more commonly known as the anomalous Hall effect (AHE) has been a remarkable tool to characterize magnetic thin films with perpendicular magnetization [1,2]. The AHE measurement technique has become highly sensitive and reliable when the magnetization is extremely small and hard to be detected by ordinary magnetization measurements such as vibrating sample magnetometer (VSM), alternating gradient magnetometer (AGM), etc. The rare-earth/transition-metal (RE-TM) family of ferrimagnet materials is one of that kind where the net magnetization ($M_S$) can be tuned from zero to a finite value by varying the composition or temperature or thickness of the alloy film [3–6]. Furthermore, the coercivity ($H_C$) in these materials follows just the opposite trend as that of $M_S$ and diverges to infinity where, $M_S$ converges to zero, known as magnetic compensation point. These properties of ferrimagnets make them an interesting candidate for a range of spintronic applications such as magneto-optical storage [7], ultrafast magnetization switching [8–11], spin-orbit torque (SOT) switching [12–17], etc. Following, we will see that how complex is the role of RE-, TM-sublattice moment and the total magnetization in governing the various transport properties. First, the SOT in these alloys, which has shown to be related to the total magnetization, was found to be diverging inversely near the compensation [14,18,19]. On the other hand, the transport properties (Hall and magneto-optic measurements) are shown to respond to the TM sublattice only because the RE 4f-electrons are located far below the Fermi level [3,4,14,20–26]. This shows the monotonic behavior of |$R_{AHE}$| with temperature (or RE concentration) from TM-rich to RE-rich cross-over [20–23]. However, in the temperature-dependent Hall resistivity measurement of $Gd_xCo_{(1-x)}$, it is observed that the Hall resistivity depends on the Gd-concentration but not on the total magnetization [27]. Moreover, in some reports, AHE is believed to be proportional to the net magnetization, and a drop in the Hall resistance at the magnetic compensation has been observed [28]. Recently, in GdFeCo, the Gd-contribution to magneto-transport is considered comparable to the FeCo-contribution, which is explained by considering the sperimagnetic nature of GdFeCo [29].

Further, unusual behavior of AHE around the compensation temperature has been observed in these alloys such as the drop of $R_{AHE}$ at a high magnetic field or the appearance of triple hysteresis loops near compensation [30–32]. Complexity in hysteresis loops, as observed in Gd-Co [33] and Gd-Fe-Bi [34] films, were thought to be originated by the magnetization gradient along the film thickness due to the composition deviation during the deposition of the film, the so-called double-layer model. However, other reports explain it differently. At a high magnetic field, a spin-flop transition, which the system does to minimize its energy by tracing the spins in the in-plane direction, is linked to these unusual AHE behaviors. So, we see there are discrepancies in the Hall contribution as well as high magnetic field AHE behavior in RE-TM alloy films, which needs to be studied for better understanding.

Here, we investigate the temperature dependence of the anomalous Hall effect in the GdFeCo sheet-film and Hall bar device. Our results of AHE on GdFeCo up to 3 T magnetic fields show the complex dependence of AHE on temperature such as, the appearance of triple hysteresis loops, anomaly in $|R_{AHE}|$ near compensation, etc., which are believed to be originated by the spin-flop transition in the system.

**MATERIALS AND METHODS**

The Gd-Fe-Co amorphous alloy film of the structure as shown in Fig. 1(a) was deposited on thermally oxidized Si-substrates using dc/rf magnetron sputtering with the sample stage spinning and orbiting simultaneously inside the chamber. The base pressure was better than 2.1 x $10^{-7}$ Torr. For GdFeCo deposition, the Gd and $Fe_{80}Co_{20}$ targets were co-sputtered at 100- and 200-watt dc power, respectively, delivering the alloy composition close to $Gd_{0.28}Fe_{0.58}Co_{0.14}$. The magnetic hysteresis of the deposited film was measured using an Alternating-Gradient Magnetometer (PMC AGM) at room temperature. After confirming the perpendicular magnetic anisotropy (PMA) in the sheet-film the Hall bar device was fabricated by electron-beam lithography and Ar ion milling. The Hall bar pattern of 5 $\mu$m width is shown in Fig. 1(b). The anomalous Hall effect (AHE) resistance measurements were carried out in a physical property measurement system (PPMS Versa Lab 3 T). The magnetic field direction is from top to bottom for a positive field value in the AHE measurements. The AHE measurements were carried out at 1 $\mu$A alternating sensing current. The present study deals with the results from the AHE resistance measurements performed on the left side independent Hall bar as shown in Fig. 1(b).

**RESULTS AND DISCUSSION**

Figure 1(c) shows the magnetic hysteresis loops of sheet-film (Fig. 1(a)) when scanning the magnetic field parallel ($H_{//}$) and perpendicular ($H\perp$) to the film plane. The hysteresis plots depict a highly square loop for the out-of-plane applied field whereas, the S-shape loop for the in-plane field, which reflects significant PMA in the sheet-film. The absence of a double loop indicates a single-phase alloy in the film. The Gd-Fe-Co film coercivity ($H_C$) and saturation magnetization ($M_S$) values are 163 Oe and 141 emu/cm$^3$, respectively. The uniaxial anisotropy ($K_U$) of the film is estimated to be ~4.5x10$^5$ erg/cm$^3$ using the relation $K_U = 1/2 H_k M_S$, where $H_k$ is the anisotropy field. $H_k$ is the field value where the in-plane magnetization starts saturating. After confirming the PMA in sheet-film we have performed AHE measurements to see the response of anomalous Hall resistance to the high magnetic field.

Figure 2(a) displays the AHE resistance loops of the sheet-film at different temperatures ranging from 50 K to 300 K. The perpendicular magnetic field up to 3 T was applied during the measurements. The sensing current in all of the AHE measurements is 1 $\mu$A and of ac in nature. The AHE resistance loops from 300 K to 260 K show the same sign of Hall coefficient and correspond to the FeCo-dominance whereas below 260 K the loops represent sign reversal assigning the Gd-dominance phase. In the present study, the Hall sign reversal governing FeCo- or Gd-sublattice dominance is subjected to the applied magnetic field direction in PPMS and the positive magnetic field refers to the perpendicular field penetrating from top to bottom of the film. The magnetic compensation temperature ($T_{comp}$) is between 240 to 260 K, where the coercivity can be seen to be diverging (Fig. 2(b)). In the 300 K AHE loop, $R_{AHE}$ remains constant up to ~ 1.7 T and decreases thereafter. Similarly, $R_{AHE}$ decreases in 280 K- and 260 K-loops after reaching the magnetic field ~ 1.3 and ~ 0.9 T, respectively. However, the opposite trend is observed after $T_{comp}$. $R_{AHE}$ starts decreasing at ~ 0.6 T for 240 K, ~ 0.7 T for 220 K, 0.9 T for 200 K, ~ 1.7 T for 150 K, and ~ 2.1 T for 100 K loops. At 50 K, $R_{AHE}$ remains constant within the applied magnetic field range (3 T). The behavior of decreasing $R_{AHE}$ after a certain applied field has also been noticed by others and is referred to as spin-flop transition and the corresponding field is called spin-flop transition field ($H_{sf}$). Generally, Gd- and FeCo-sublattice moments are collinear and the dominant sublattice moment (FeCo) is along the field direction. However, after reaching a certain field the Gd-sublattice moment pulled towards magnetic field direction by the angle ($\alpha$) from the antiferromagnetic axis, and FeCo, being antiferromagnetically coupled to Gd, also moves away from the field direction by the angle ($\beta$) as shown in Fig. 2(c). This applied field where both sublattices are no longer collinear is called the spin-flop transition field ($H_{sf}$). At a somewhat larger field, the spins are flopped completely and align along the x-axis, this field is called the spin-flopped field ($H_{sfo}$). However, at a very large magnetic field (>>3 T) both sublattice moments can finally become parallel to the field direction, such field is called spin-flip transition field ($H_{sfi}$) [35,36]. Figure 2(b) shows the spin-flop field variation as a function of temperature. $H_{sf}$ is smallest when the temperature is approaching the $T_{comp}$, but it keeps increasing on going away from the $T_{comp}$. In the present work, we report a very small spin-flop transition field of 0.6 T at 240 K. Moreover, when comparing the change in $H_{sf}$ above and below the $T_{comp}$, it changes faster above the $T_{comp}$.

The total energy of the system can be written as [37]

$$E = -HM_S \cos\theta - 2\pi M_S^2 \sin^2\theta + K_U \sin^2\theta - \lambda M_{TM} M_{RE} \cos(\alpha+\beta),$$

where $M_S$ is net magnetization ($M_{TM}-M_{RE}$) of the film, $\theta$ is the angle of $M_S$ from the film normal, $\lambda$ is the exchange coupling constant, and $\alpha+\beta$ is the net canting angle between the two sublattices. The four terms are external field energy, demagnetization energy, uniaxial anisotropy energy, and exchange coupling energy density, respectively.

The expression for $\lambda$ is given by [37]

$$\lambda = \frac{2(Z+1)|J_{RE-TM}|}{N g_{RE} g_{TM} \mu_B^2}$$

here, $J_{RE-TM}$ is the exchange energy integral of RE-TM pair, $z$ is coordination number, $N$ is total atomic number density, $g$ is gyromagnetic factor and $\mu_B$ is Bohr magneton.

The flopping of sublattice magnetization can also be understood from the exchange coupling constant ($\lambda$), which characterizes the canting strength of RE-TM subnetworks [5,37]. It has been found in Ref. [5] that the $\lambda$ value is significantly smaller (~500) in RE-TM alloys with Gd in comparison to others like Tb (~$10^4$) and Dy (~$10^3$), which also reflects in the small value of spin-flop field for GdFeCo. The external field competes with exchange coupling energy to flop the sublattice magnetization towards the in-plane axis at $H_{sfo}$ and flips both sublattices to the same direction (perpendicular axis) at $H_{sfi}$.

From the AHE loops, we have also calculated the difference between the up and down state $R_{AHE}$ i.e., $\Delta R_{AHE}$. Figure 3 shows the temperature dependence of $\Delta R_{AHE}$ obtained from the 3 T-AHE loops (Fig. 2) at remnant and near saturation (3 T) states. The latter includes spin-flop contribution. In the remnant $\Delta R_{AHE}$ curve, before compensation, the $|\Delta R_{AHE}|$ decreases with increasing the temperature with a sudden drop near compensation (240 K). Similarly, after compensation with obvious sign reversal, the $\Delta R_{AHE}$ magnitude decreases slowly with an increase in the temperature. In the case of 3 T $\Delta R_{AHE}$ curve, the $\Delta R_{AHE}$ continuously drops with an increase in the temperature irrespective of crossing the compensation point. At 280 K, it decreases to an infinitesimally small magnitude of 0.19 Ω with holding a negative sign (opposite to sign reversal). This anomalous sign convention suggests a strong spin-flop transition. The large decrease in anomalous Hall resistance indicates the orientation of sublattice moments towards the in-plane axis [34]. However, the sublattice moments are still antiferromagnetically coupled and not acquired the spin-flip transition. The $\Delta R_{AHE}$ increases to 2.46 Ω at 300 K. The two curves overlap each other at far below the compensation temperature (50 and 100 K) where a significantly larger field is required to flop the spin. Above room temperature, where the thermal effects are dominant, the remnant state $R_{AHE}$ drops when approaching the curie temperature, where both curves should be meeting again.

To further see how the spin-flop transition behaves in microchannels of the film, we have fabricated the Hall bar device and performed the AHE loop measurements at different temperatures. Figure 4 shows the AHE loops of the GdFeCo Hall bar device at different temperatures. From AHE loops it appears that the $T_{comp}$ is close to 118 K, where sign reversal takes place. The different compensation points for the sheet film and Hall bar device are possibly due to the fabrication process and the average homogeneity of the sheet film. The coercivity diverges on increasing the temperature towards the compensation. However, anomalies in the coercivity can be seen just above the compensation as shown in Fig. 4(b). Above $T_{comp}$, one can see the triple loops at 130 K, 120 K, and 119 K which could be the reason behind the unusual dip in the coercivity just above the compensation. The presence of triple loops around the compensation has also been observed by other researchers in the GdFeCo [38], TbFeCo [30], and other RE-TM films [33,34,39,40]. There are inconsistencies in the occurrence of triple loops around the compensation. In some systems, they appear below $T_{comp}$ but in others above $T_{comp}$. For instance, in pristine GdFeCo and TbFeCo, triple loops occur just above the $T_{comp}$, on the other hand, in Ta-capped TbFeCo [30] they appear just below the $T_{comp}$. In the present work i.e., Hf-seeded GdFeCo, they appear just above the $T_{comp}$. The origin of triple loops has been explained by considering the magnetic anisotropies of RE and TM elements in the film [30,32]. Further, the occurrence of triple loops below or above compensation has also been demystified by considering the dominance of the FeCo or Gd/Tb-anisotropy. In pristine GdFeCo or TbFeCo alloy usually, the RE anisotropy prevails over the TM which results in collinear to non-collinear first-order phase transition toward the higher temperature and therefore the appearance of the triple loops [32]. In the case of Ta-capped ferrimagnetic film, the FeCo anisotropy increases due to the surface anisotropy introduced by Ta, which governs triple loops below the compensation temperature [30]. It is noticeable that here we use Hf as an adjacent layer, however, in Ref. [30], the bottom layer, as well as the top layer, was Ta. We suggest that the Hf couldn't introduce surface anisotropy to FeCo that is why our case is similar to the pristine GdFeCo or TbFeCo, where the Gd or Tb anisotropy prevails and results in the appearance of triple loops just above the compensation.

The difference in the AHE behavior of sheet film and the device can be due to the following reasons: The RE-TM alloys are sensitive to heating during the spin coating in the fabrication process because of that the sample properties (compensation point, coercivity, magnetization, etc.) change drastically. Further, at low temperatures, the increased rare-earth anisotropy could be the possible reason behind the difference. Another reason might be the homogeneity of the film under inspection. Hall cross (5 µm width) is much smaller than the sheet film (5 mm x 5 mm), therefore, an average behavior of AHE can be seen in the sheet film.

A frequently arising question in the ferrimagnet system is the contribution of RE or TM moments in AHE. To understand this, we have carried out temperature-dependent AHE resistance measurement in absence of a magnetic field. Figure 5(a) shows the temperature-dependent $R_{AHE}$ scan during cooling the film without the magnetic field. In the two experiments, magnetization was set in the up ($M_z\uparrow$) and down ($M_z\downarrow$) state, respectively, before running the measurement. Since we start measurement at 300 K where the ferrimagnet is FeCo dominant, therefore, the $M_z\uparrow$ means FeCo up and Gd down. However, after crossing the $T_{comp}$, the dominance of sublattice moments reverse but as there is no applied field so the Hall sign remains unchanged. However, there is no change in the magnitude of AHE when the net magnetization is close to zero. This supports the argument that the FeCo solely contributes to the AHE [21,41]. Furthermore, for both $M_z\uparrow$ and $M_z\downarrow$ states, the magnitude of Hall resistance increases gradually on cooling. It can be seen that the $R_{AHE}$ increases on cooling the film. This monotonic increase in the AHE on cooling the device is similar to the FeCo magnetic moment response to the cooling [42].

To know how the AHE will respond when a magnetic field is applied, we performed the same measurements at the 3 T applied field. Here, we have a special situation of spin-flopping so the 3 T AHE becomes important to know. Figure 5(b) shows the two $R_{AHE}$ curves, one at -3 T perpendicular magnetic field ($M_z\uparrow$) while cooling the sample from 300 K to 50 K and the other at 3 T ($M_z\downarrow$) while warming the sample. Both the curves are symmetric to each other. Unlike zero-field measurements here we see sign reversal across the compensation. The two curves cross each other at compensation, where Gd and FeCo sublattices tend to cancel each other. Therefore, Fig. 5(b) suggests that the AHE is coming from the net magnetization which is close to zero near the compensation, which is unlike the case in Ref. [41] but similar to some other reports [28]. The conclusions coming from these two figures are contradicting each other. However, this can be explained on behalf of spin-flop behavior in the particular region where the contradiction arises. Due to the low field spin-flopping around compensation, the FeCo moments flops towards the film plane in the presence of a magnetic field, consequently dropping the AHE signal. The $R_{AHE}$ signals approach a similar value in these two figures, where the spin-flopping is small. However, further understanding of the particular phenomena needs to be explored in the future.

## CONCLUSIONS

In summary, we investigated the Hf/GdFeCo/MgO sheet film and Hall bar device for the temperature dependence of anomalous Hall effect up to 3 T magnetic field. Both the sheet film and device have shown compensation around 240 K and 118 K, respectively. The triple hysteresis loops have been observed in the Hall bar device above the compensation where the coercivity is seen to be falling unusually. The zero-field temperature-dependent AHE results suggest only FeCo contribution to the Hall signal. The temperature scan of $R_{AHE}$ with magnetic field suggests the Hall signal considers both Gd and FeCo moments. The contradicting response of the latter case might be due to the low field spin-flopping of FeCo moments in the device.

## ACKNOWLEDGMENT

This work was supported by the Ministry of Science and Technology (MOST) Taiwan ROC (Grant No. MOST 109-2112-M-224-001-MY2). The author RCB would like to acknowledge the National Yunlin University of Science and Technology, Taiwan ROC for funding post-doctoral research through University Postdoc Fellowship.

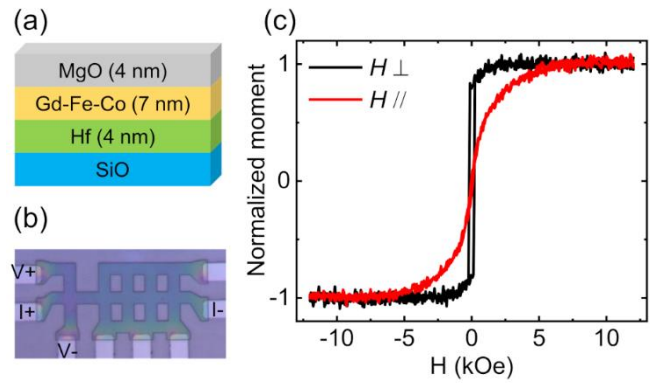

**Fig. 1** (a) Schematic of film structure, (b) optical image of Hall bar pattern, and (c) magnetic hysteresis loops of GdFeCo sheet-film measured at room temperature.

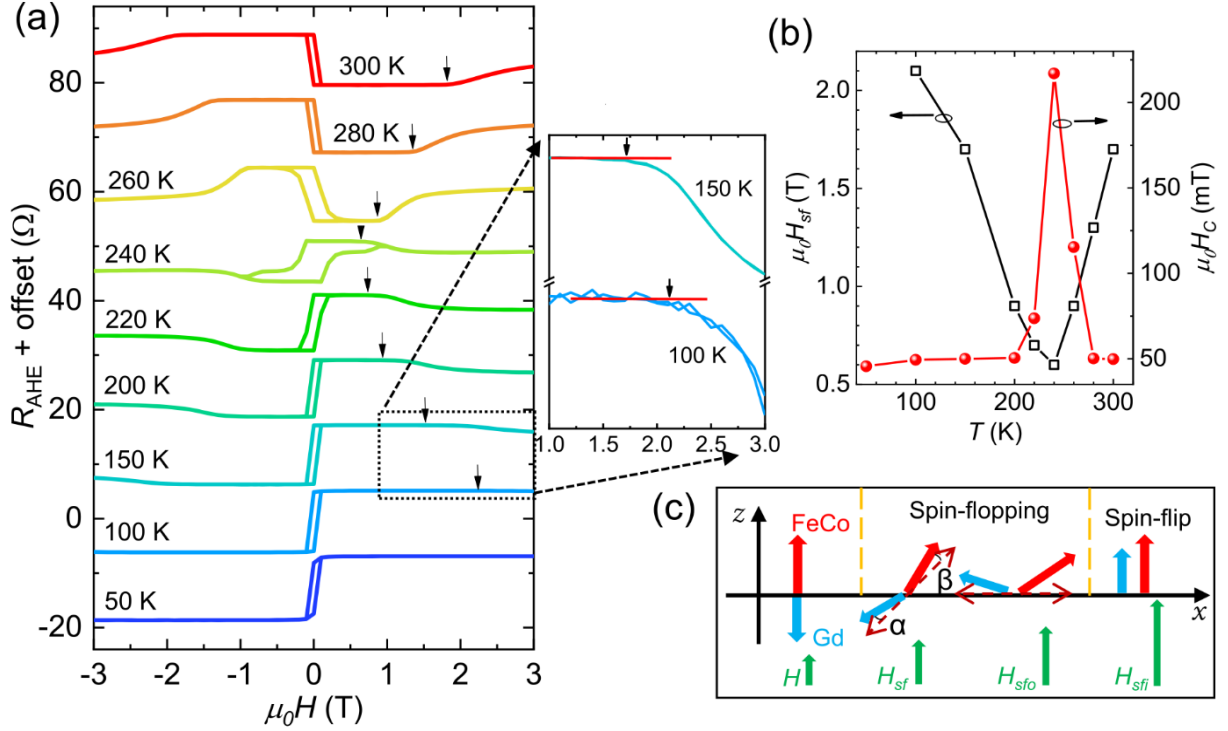

**Fig. 2** (a) AHE resistance loops of GdFeCo sheet-film at different temperatures. The 50 K-loop is virgin but the other loops are shifted vertically upward with random offsets. The magnified view shows clear drop in the $R_{AHE}$ at spin-flop field for the 100 K and 150 K loops. (b) Variation of $H_C$ and spin-flop transition field ($H_{sf}$) as a function of temperature. The $H_C$ peaks around the compensation temperature, whereas the $H_{sf}$ falls around the same. $H_{sf}$ is shown by black arrows in the AHE resistance loops. (c) Spin-flop model for FeCo dominant GdFeCo (adapted from Ref. [35,36]): On increasing the magnetic field ($H$) beyond the remnant state, the FeCo and Gd sublattice moments start flopping towards the in-plane axis (i.e. leading to $H_{sf}$ and then $H_{sfo}$) before flipping to the applied field direction at the comparatively large field ($H_{sfi}$).

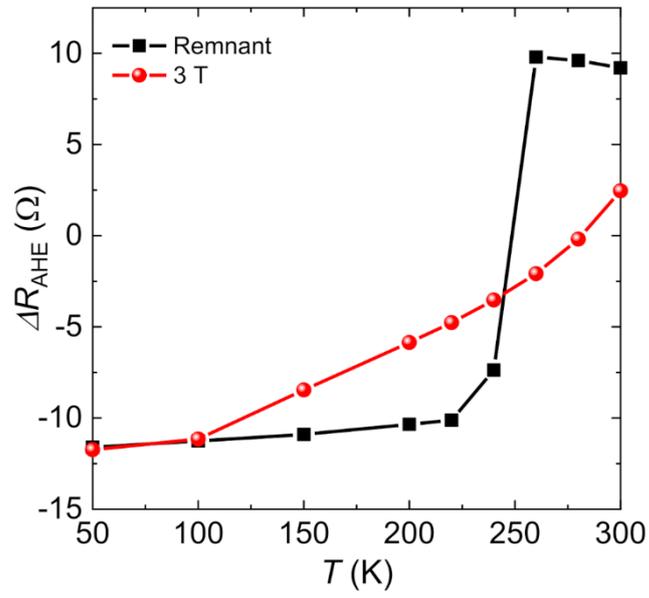

**Fig. 3** Temperature dependence of $\mathit{\Delta R}_{AHE}$ obtained from the 3 T AHE resistance loops (Fig. 2) at remnant and near-saturation (3 T) states.

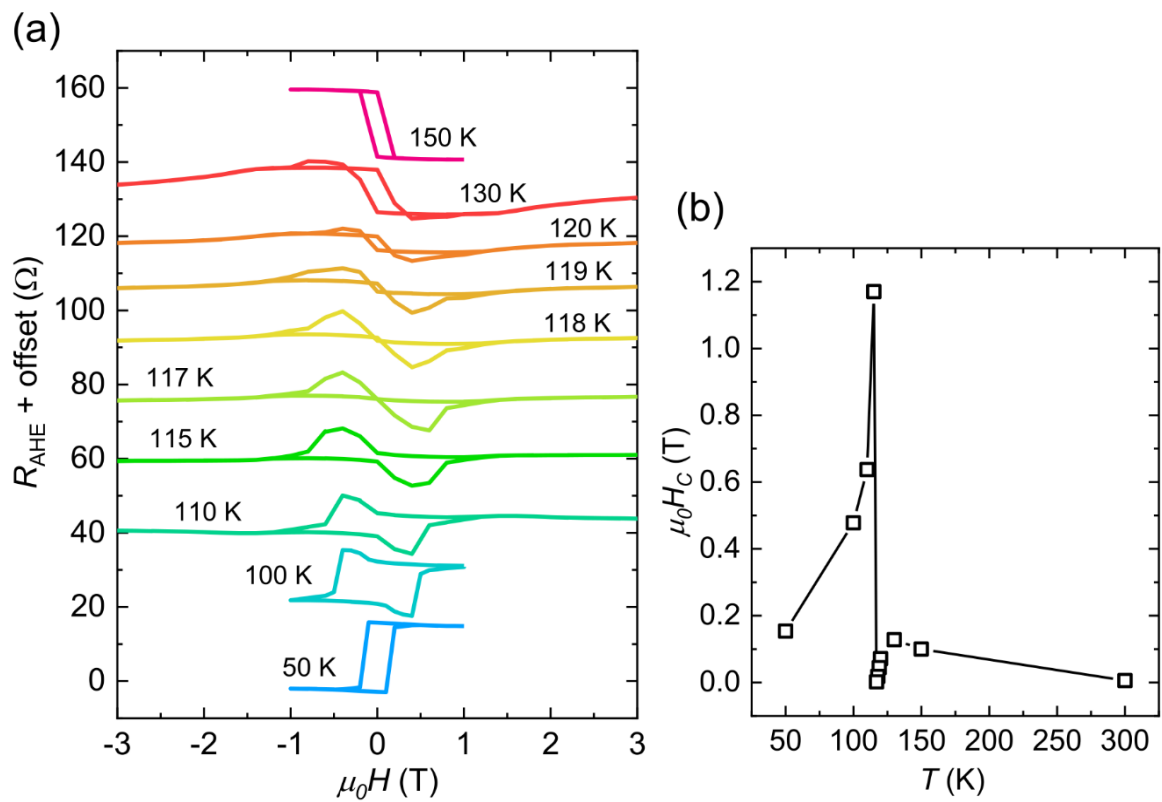

**Fig. 4** (a) AHE resistance loops at different temperatures for the GdFeCo Hall bar device, and (b) the variation of coercivity as a function of temperature derived from the AHE resistance loops.

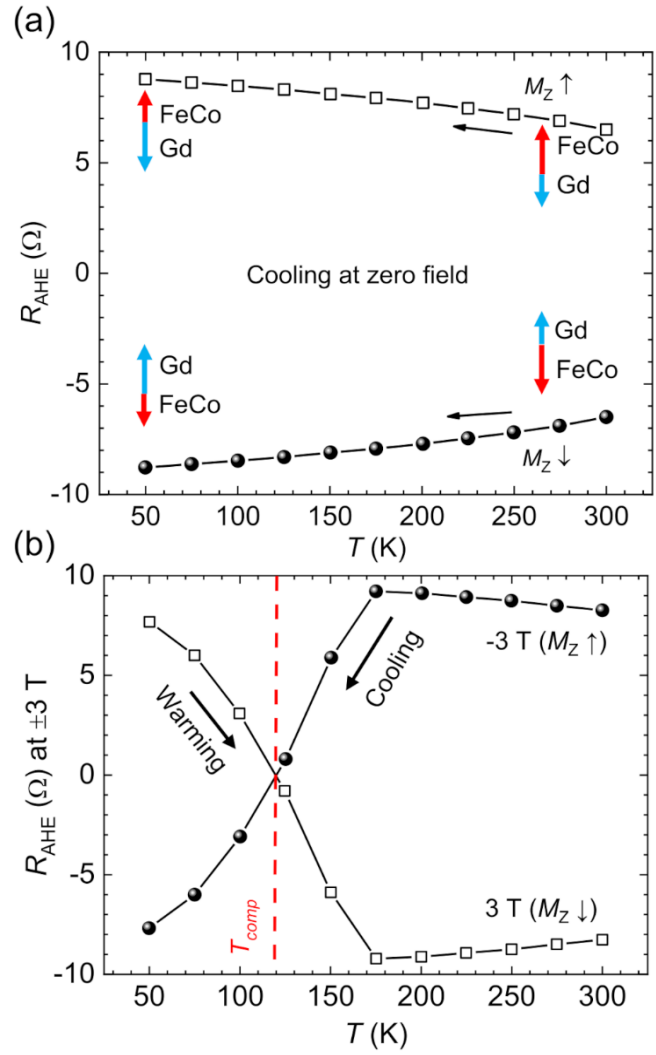

**Fig. 5** (a) Temperature-dependent $R_{AHE}$ scan while cooling the Hall bar device without external magnetic field. The device was set in up ($M_Z\uparrow$) and down ($M_Z\downarrow$) magnetization states before scanning the resistance. (b) Temperature-dependent $R_{AHE}$ scan while cooling and warming at 3 T magnetic field perpendicular to the film plane. The instrumental offset has been removed from the data plots in (a) and (b).